\documentclass[preprint,aps,showpacs,superscriptaddress,preprintnumbers,amsmath,amssymb]{revtex4}

\usepackage{graphicx}
\usepackage{dcolumn}
\usepackage{bm}

\newcommand{\rbo}{\protect\mbox{C($^{17}$C,$^{15}$B+$n$)X}}
\newcommand{\rbk}{\protect\mbox{C($^{17}$C,$^{15}$B)X}}

\newcommand{\rhe}{\protect\mbox{C($^{14}$B,$^6$He+$n$)X}}

\newcommand{\cbo}{\protect\mbox{$^{15}$B-$n$}}

\newcommand{\etal}{\textit{et al.}}

\newcommand{\nima}{Nucl. Inst. Meth. A}
\newcommand{\plb}{Phys. Lett. B}

\newcommand{\epja}{Eur. Phys. J. A}

\begin{document}


\title{Single-Proton Removal Reaction Study of $^{16}$B}

\author{J.-L.~Lecouey}
\affiliation{LPC-Caen, ENSICAEN, Universit\'e de Caen et IN2P3-CNRS, 14050 Caen Cedex, France}
\author{N.A.~Orr}
\affiliation{LPC-Caen, ENSICAEN, Universit\'e de Caen et IN2P3-CNRS, 14050 Caen Cedex, France}
\author{F.M.~Marqu\'es}
\affiliation{LPC-Caen, ENSICAEN, Universit\'e de Caen et IN2P3-CNRS, 14050 Caen Cedex, France}
\author{N.L.~Achouri}
\affiliation{LPC-Caen, ENSICAEN, Universit\'e de Caen et IN2P3-CNRS, 14050 Caen Cedex, France}
\author{J.-C. Ang\'elique}
\altaffiliation{Present address: LPSC, Grenoble, France}
\affiliation{LPC-Caen, ENSICAEN, Universit\'e de Caen et IN2P3-CNRS, 14050 Caen Cedex, France}
\author{B.A.~Brown}
\affiliation{NSCL, Michigan State University, East Lansing, MI 48824-1321, USA}
\author{W.N.~Catford}
\affiliation{Department of Physics, University of Surrey, Guildford GU2~7XH, UK}
\author{N.M.~Clarke}
\affiliation{School of Physics and Astronomy, University of Birmingham, Birmingham B15~2TT, UK}
\author{M.~Freer}
\affiliation{School of Physics and Astronomy, University of Birmingham, Birmingham B15~2TT, UK}
\author{B.R.~Fulton}
\affiliation{Department of Physics, University of York, York YO10~5DD, UK}
\author{S.~Gr\'evy}
\altaffiliation{Present address: GANIL, Caen, France}
\affiliation{LPC-Caen, ENSICAEN, Universit\'e de Caen et IN2P3-CNRS, 14050 Caen Cedex, France}
\author{F.~Hanappe}
\affiliation{PNTPM, CP-229, Universit\'e Libre de Bruxelles, B-1050 Brussels, Belgium}
\author{K.L.~Jones}
\altaffiliation{Present address: Dept of Physics and Astronomy, Rutgers University, New Jersey, USA}
\affiliation{Department of Physics, University of Surrey, Guildford GU2~7XH, UK}
\author{M.~Labiche}
\altaffiliation{Present address: STFC Daresbury Laboratory, Warrington, UK}
\affiliation{LPC-Caen, ENSICAEN, Universit\'e de Caen et IN2P3-CNRS, 14050 Caen Cedex, France}
\author{R.C.~Lemmon}
\altaffiliation{STFC Daresbury Laboratory, Warrington, UK}
\affiliation{Department of Physics, University of Surrey, Guildford GU2~7XH, UK}
\author{A.~Ninane}
\affiliation{Slashdev Integrated Systems Solutions, Rue de la Rochette, 26, B-5030 Gembloux, Belgium}
\author{E.~Sauvan}
\altaffiliation{Present address: CPPM, Marseille, France}
\affiliation{LPC-Caen, ENSICAEN, Universit\'e de Caen et IN2P3-CNRS, 14050 Caen Cedex, France}
\author{K.M.~Spohr}
\affiliation{School of Engineering and Science, University of Paisley, Paisley PA1 2BE, Scotland}
\author{L.~Stuttg\'e}
\affiliation{IPHC, Universit\'e Louis Pasteur et IN2P3-CNRS, 67037 Strasbourg Cedex 02, France}


\begin{abstract}

The low-lying level structure of the unbound system $^{16}$B has been investigated via
single-proton removal from a 35~MeV/nucleon $^{17}$C beam.
The coincident detection of the beam velocity $^{15}$B fragment and neutron allowed the 
relative energy of the in-flight decay of 
$^{16}$B to be reconstructed.  The resulting spectrum exhibited a 
narrow peak some 85 keV above threshold.  It is argued that this feature corresponds to a very 
narrow ($\Gamma \ll $100~keV) resonance, or an unresolved multiplet, with a
dominant $\pi (p_{3/2})^{-1} \otimes  \nu (d_{5/2}^3)_{J=3/2^+}$ + 
$\pi (p_{3/2})^{-1} \otimes  \nu (d_{5/2}^2,s_{1/2})_{J=3/2^+}$ configuration which decays 
by $d$-wave neutron emission.

\end{abstract}


\maketitle

Amongst the light exotic nuclei, the two-neutron halo systems are arguably the most intriguing.
Apart from the halo character, these nuclei exhibit the so-called ``Borromean'' property
whereby, in a three-body description, none of 
the constituent two-body subsystems --- $n$-$n$ and core-$n$ -- are bound \cite{Han95}. Modelling
such systems thus requires knowledge of the core-$n$ interaction.
Given that measurements of neutron scattering on the core nucleus are in practice 
impossible \footnote{Except 
for the case of 
$^6$He and the $\alpha$-n interaction.}, the interaction must be derived from the structure of the
corresponding unbound nucleus --- $^{10}$Li, for example, in the case of $^{11}$Li \cite{Tho94}.

Currently the heaviest established two-neutron halo system is $^{17}$B \cite{Yam04}.
Analyses of both the $^{15}$B fragment momentum distribution following the breakup of
a $^{17}$B beam \cite{Suz02} and the total reaction cross section on a carbon target \cite{Yam04} 
indicate that the halo neutrons wavefunction contains roughly equal admixtures of $\nu 1d_{5/2}^2$
and $\nu 2s_{1/2}^2$ configurations.  The recent observation of the gamma-ray de-excitation
of the two bound excited states of $^{15}$B \cite{Sta04} in the dissociation of $^{17}$B suggests that core excitations may also play a role \cite{Kan05,Kon05}.

Very little, however, is known about $^{16}$B.  Its unbound nature was
most recently confirmed by Kryger \etal\ in an investigation of the reaction products from the breakup of 
$^{17}$C where an upper limit on the lifetime of some 191 ps was derived \cite{Kry96}.
Evidence for a very low-lying state, some 40~keV above the \cbo\ threshold, together with indications for a higher lying level
2.40~MeV above threshold was found in a heavy-ion multi-nucleon transfer reaction study \cite{Kal00}.  
Whilst benefiting from the advantages of the high intensities of stable beams, 
such reactions suffer from very low yields (cross sections of order $\mu$b/sr and 
thin targets -- $\sim$1~mg/cm$^2$),
a complex reaction mechanism and hence selectivity, often coupled with significant backgrounds as illustrated by Ref.~\cite{Kal00}.

As originally demonstrated by the investigation of $^{10}$Li by Zinser \etal\ \cite{Zin95}, the 
few-nucleon breakup of high-energy radioactive beams can be employed to populate, and study through the
fragment--neutron final-state interaction (FSI), unbound nuclei.  In addition to benefiting from
high cross sections (typically 10-100's~mb), the high energies result in the strong forward focussing
of the reaction products (increasing the effective detection acceptances) and permit the use of 
very thick targets ($\sim$100's~mg/cm$^2$).  Consequently measurements with beam intensities as low
as $\sim$100~pps are feasible.  Importantly, as discussed by Zinser \etal\ within the context of the sudden approximation,
the stripping at high energy of
one or more protons (for example) will not perturb the configuration of the neutrons.
As such the choice of a projectile with a well known structure should
permit the structure of the final-states of the unbound system to be inferred.
Indeed, the use of a $^{11}$Be beam with a predominately $\nu s_{1/2}$ configuration 
allowed the threshold $s$-wave structure of $^{10}$Li to be probed \cite{Zin95,Che01}.
More recently, two-proton removal from $^{11}$Be has been employed to explore the 
structure of $^9$He \cite{Che01}.
 
In this spirit we report here on an investigation of the low-lying level structure of 
$^{16}$B using single-proton
removal from a high-energy beam of $^{17}$C.  As discussed below, the shell structure of $^{17}$C 
has been established by a number of recent experiments \cite{Sau00,Mad01,Oga02,Sau04,Elk05}.

The $^{17}$C beam of mean energy 35~MeV/nucleon was
produced via the reaction on a Be target of a 63~MeV/nucleon $^{18}$O primary beam supplied
by the {\sc ganil} cyclotron facility.
The beam velocity reaction products were analysed and purified 
using the {\sc lise}3 fragment separator \cite{Ann87}.
The resulting secondary beam composition was 98\% $^{17}$C with an average beam intensity of $\sim $7000~pps.  
A time-of-flight measurement performed using the beam tracking detectors (see below) and a
parallel-plate avalanche counter ({\sc PPAC}) located some 24~m upstream of the secondary reaction
target allowed the $^{17}$C ions to be separated in the offline analysis from the remaining 
contaminants.  The energy spread of the $^{17}$C beam, as defined by the settings of 
the {\sc lise} spectrometer, was 2.5\% ($\Delta E/E$).  The effect on
the relative energy between the fragment and neutron from the in-flight decay of an unbound system, such as
$^{16}$B, is negligible compared to the overall resolution (see below) and no event-by-event
correction based on the measured beam energies was required. 

The secondary beam was delivered using the separator onto a 95~mg/cm$^2$ $^{nat}$C target.
Owing to the relatively poor optical qualities of the secondary beam (beam spot size $\sim$10~mm diameter),
two position-sensitive {\sc ppac}'s located just upstream of the secondary reaction target and
seperated by 60~cm were
used for tracking.  The impact point of the beam on target was thus reconstructed event-by-event with a 
resolution of $\sim$1.5~mm ({\sc fwhm}).
The charged fragments from the reactions were detected and identified 
using a Si-Si-CsI telescope centred at 0$^\circ$ and located 11.3 cm
downstream of the target. 
The two 500~$\mu$m thick silicon detectors, each comprising 16 resistive strips, were mounted
such that the strips of the first detector provided for a measurement of position in the horizontal 
direction whilst those of the second detector the vertical position.  The impact point along each
strip was determined with a resolution of 1~mm ({\sc fwhm}).
In addition to the energy-loss measurements provided by the silicon detectors, the residual energy
of each fragment was determined from the signals derived from the 5$\times$5~cm$^2$, 
2.5 cm thick CsI 
crystal.  The measurement of the total energy deposited in the telescope was calibrated using
a``cocktail'' beam, which included $^{15}$B, and for which the energy spread was
0.05\%.  Runs were made over a range of rigidity settings of the {\sc lise} spectrometer such that
the $^{15}$B calibration points covered the range of energies expected from the in-flight decay 
of $^{16}$B. 
The total energy resolution
of the telescope was determined to be 1.2\% ({\sc fwhm}).
 In addition to the measurements of breakup on the C target, data was also acquired
with the target removed so as to ascertain the contribution arising from reactions in the
telescope.  As the reaction of interest -- \rbo\ -- is a charge changing one, this contribution
was found to be negligible as expected \cite{Mar96}.

The neutrons were detected using 97 liquid scintillator elements of the {\sc demon} array \cite{Til95}. The modules, each of which has an intrinsic detection efficiency of $\sim$35\% at 35 MeV, were 
arranged in a staggered two-wall type configuration
covering polar angles up to 39$^\circ$ in the laboratory frame \cite{Lec03}.  This arrangement
provided for a reasonable detection efficiency for the \cbo\ reaction out to $\sim$5~MeV relative energy,
whilst retaining a good energy resolution (see below).  The neutron energy was derived from the
time
of flight measured between the {\sc ppac} located closest to the target and
{\sc demon}.  The final energy resolution was 5\% ({\sc fwhm}).  The neutron energy
spectrum exhibited, in addition to the beam velocity neutrons produced in the projectile breakup, a low-energy 
component arising from neutrons evaporated from the 
target.  Such events were removed in the off-line analysis by imposing a low-energy threshold of 
13 MeV \cite{Lec03}.

\begin{figure}[htb]
   \begin{center}
    \includegraphics[width=12cm]{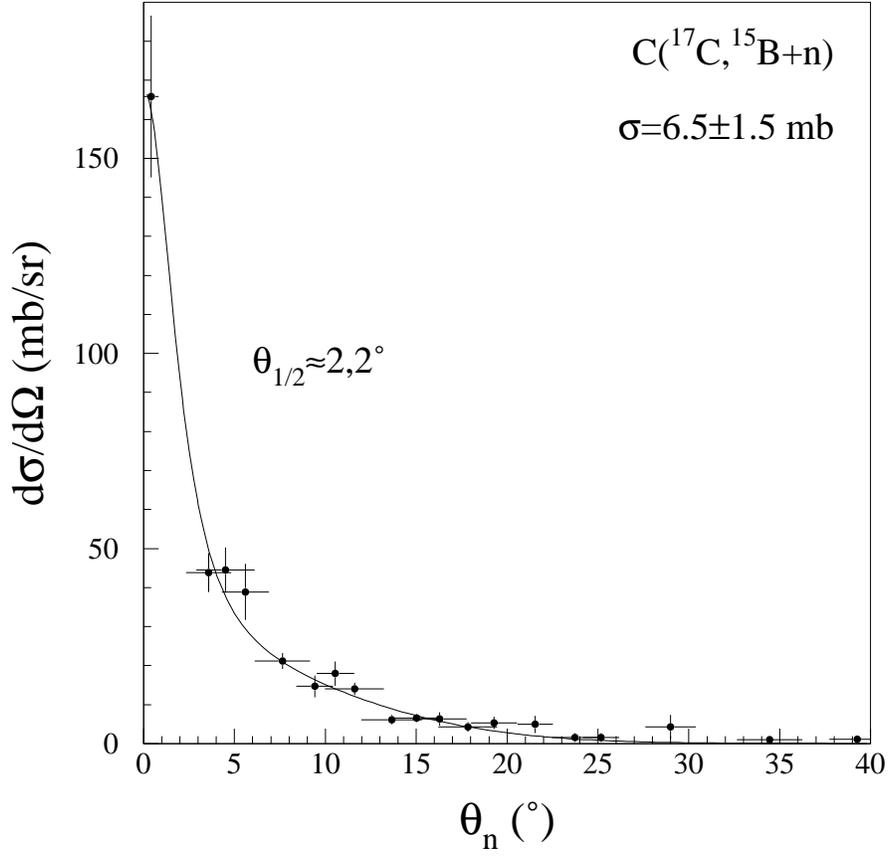}
   \end{center}
   \caption{\small \rm Single-neutron angular distribution for the reaction \rbo\ at
35~MeV/nucleon.  The solid line is an adjustment 
to the data using a Lorentzian plus Gaussian lineshape.}
   \label{fig:angdis}
\end{figure}

Turning to the results, the most easily extracted observable is the single-neutron angular distribution in
coincidence with the $^{15}$B fragments.  As may be seen in Figure \ref{fig:angdis}, the angular distribution is
very forward focussed ($\theta_{1/2} \approx 2^\circ$), indicating that the in-flight decay is dominated by
events with low relative energies.  The angle integrated cross section  
is 6.0$\pm$1.5~mb, in good agreement
with the value of is 4.4$\pm$0.3~mb obtained by Krieger \etal\ at a
somewhat higher beam energy (52~MeV/nucleon) for the inclusive channel \rbk\ \cite{Kry96}.

\begin{figure}[htb]
   \begin{center}
      \includegraphics[width=12cm]{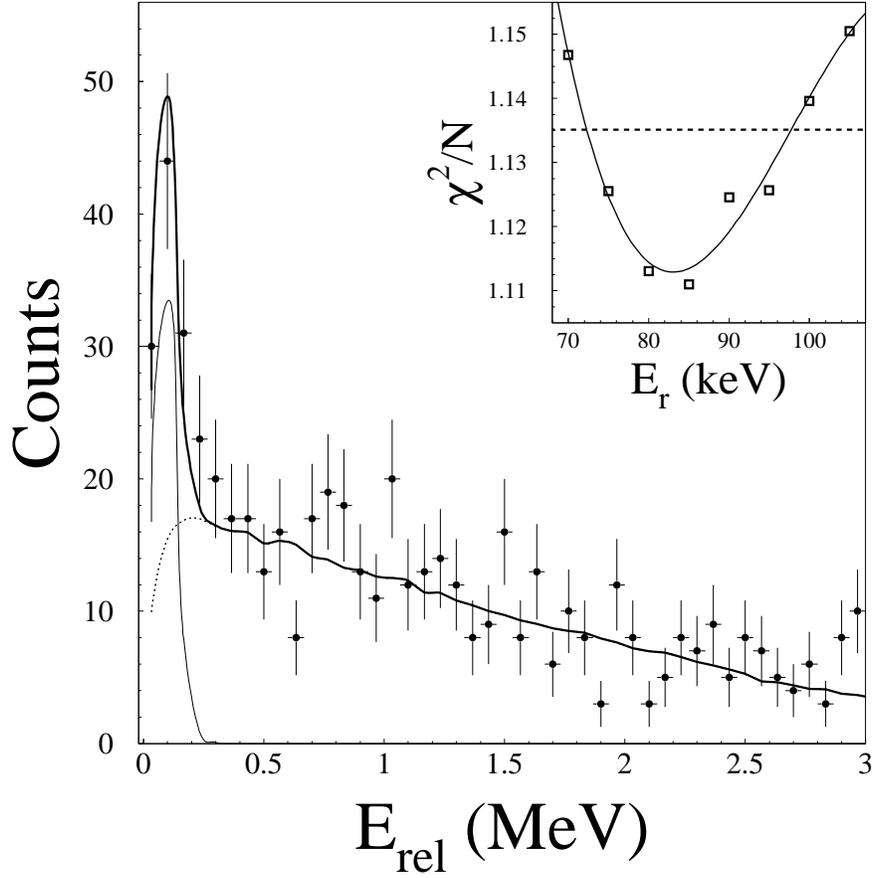}
      \caption {\small \rm  Reconstructed \cbo\ relative energy spectrum. The thick solid line corresponds
          to the best adjustment to the data for a very narrow $d$-wave resonance 
($E_r$=85$\pm$15~keV, $\Gamma$ = 0.5 keV) folded with the simulated experimental response function 
(thin solid line) plus a broad uncorrelated distribution 
obtained by event-mixing (dotted line) -- see text. The insert shows the reduced $\chi^2$ as a 
function of $E_r$, where the horizontal line deliniates $\chi^2$+1.}
        \label{fig:narrow16b}
   \end{center}
\end{figure}

The reconstructed \cbo\ relative-energy spectrum is displayed in Figure \ref{fig:narrow16b}.  As expected,
significant strength resides close to the decay threshold, in particular in the form of 
a very narrow structure ({\sc fwhm} $\approx$ 100 keV) at $\sim$100~keV.
It should be pointed out that the measured relative energy may be directly identified with the
energy in $^{16}$B provided that the $^{15}$B fragment is in the ground state.  As noted earlier,
$^{15}$B was recently found to possess two bound excited states (E$_x$=1.33 and 2.73~MeV) \cite{Sta04}.  
Given the relatively limited yield (655 events) in the \rbo\ channel (Figure \ref{fig:narrow16b}), a triple 
coincidence measurement including gamma-ray detection was precluded.  Structural considerations 
(see below) coupled to the relative high-energies of the bound excited states in $^{15}$B, lead to 
the conclusion, however,
that the feature observed does not arise from the decay of a high-lying level in $^{16}$B to 
a bound excited state in $^{15}$B.

\begin{figure}[htb]
   \begin{center}
     \includegraphics[width=12cm]{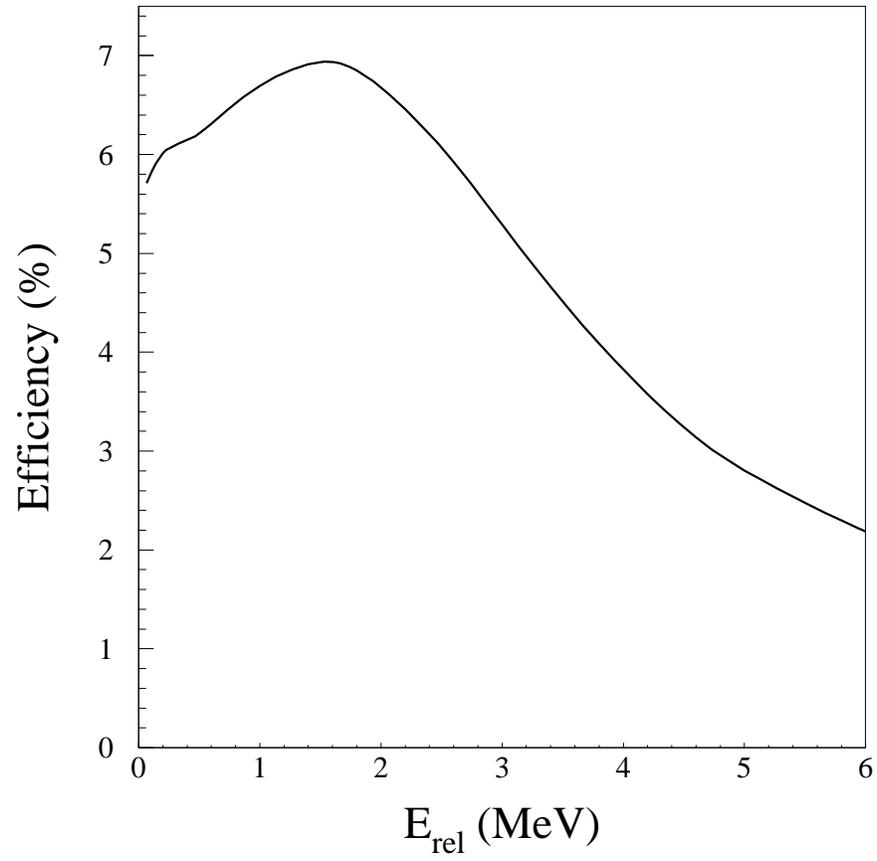}
   \end{center}
   \caption{\small \rm Simulated detection efficiency for the \cbo\ pair as a function of relative energy (see text for details).}
   \label{fig:effdecay}
\end{figure}

Before proceeding any further in the interpretation, the effects of the experimental setup
must be examined.  The experimental response functions were generated using a
{\sc geant} \cite{Gea87} based simulation code \cite{Lab99,Lab01}.  The predicted efficency for detecting 
a \cbo\ pair is shown in Figure \ref{fig:effdecay}.  Importantly, the response is a smooth function
of relative energy exhibiting no features which could mimic a sharp resonance-like state.  The gradual
fall off
from a maximum of some 7\% at around 1.5~MeV is in line with simple geometrical considerations based
on the angular coverage of the neutron array (the efficiency for the detection of the $^{15}$B fragments
is essentially 100\% ). 
The resolution ({\sc fwhm}) in the reconstructed relative energy, which is dominated by the 
finite angular size of the individual {\sc DEMON} modules, was determined to 
vary as 0.320$\cdot \sqrt{E_{rel}}$~(MeV), with a resolution of 100~keV at 100~keV decay energy.
This suggests that the width of the feature observed in the decay spectrum (Figure \ref{fig:narrow16b})
is dominated by the experimental resolution.
As a check on the simulations, comparison was made
between the $^7$He decay spectrum -- the ground state resonance of which is well established --
reconstructed from data acquired in the \rhe\ reaction at 41~MeV/nucleon and that predicted by the Monte Carlo calculations.
As discussed elsewhere the agreement was very good \cite{Lec03,Lec07}.
Cross-talk events, whereby scattering resulted in two detectors
firing, were discarded by analysing only single-neutron events.  The probability of cross-talk occuring
was well reproduced by the simulations.
Finally, we note that rate of events for which
a neutron was first scattered without detection by a module (or non-active element in the setup) and then detected in another module was predicted to be less than $\sim $5\% of the
total number of events and did not introduce noticeable distortions in the reconstructed relative 
energy spectrum.    

In addition to the sharp resonance-like peak, the reconstructed \cbo\ relative-energy spectrum
exhibits a very broad underlying distribution which slowly decays in intensity with increasing 
energy (Figure \ref{fig:narrow16b}).  This distribution is interpreted as arising from
the population of the non-resonant continuum -- that is $^{15}$B-n events for which no
significant FSI occurs \footnote{Given the limited resolution at high relative energies, the presence of
weakly populated high-lying resonances cannot be discounted.}.  Such uncorrelated events may be generated from the experimentally 
measured \cbo\ pairs via event mixing.  In order to avoid the effects of possible residual correlations 
\cite{Zaj84} an iterative technique was applied \cite{Mar00}.  The distribution so generated and
normalised to the data at high relative energy is
shown in Figure \ref{fig:narrow16b} (dashed line).  The agreement is very good and reinforces the notion 
that the resonance-like peak is not an artifact of the experimental setup \footnote{The uncorrelated distribution so generated includes the effects of the experimental response function.}.

In order to interpret the data, the formalism developed in Refs \cite{Zin95,Che01} has been followed. 
Briefly, within the sudden approximation, the lineshape of the relative
energy distribution is derived from the overlap of the initial bound-state wave function, describing the relative motion between the valence neutron and core
of the projectile, and the unbound final-state wave function describing the fragment-neutron interaction. Here the wave functions were obtained using standard Woods-Saxon
potentials ($r_0$=1.25~fm, $a_\nu$=0.6~fm) adjusted to reproduce the neutron separation energy for the intial state and the resonance energy for the final state.
As described in Ref. \cite{Lec03}, for a narrow resonance ($\ell_n >$~0)
this approach results in a lineshape identical to a Breit-Wigner distribution incorporating the appropriate $\ell_n$-dependent penetrability.

As noted in the introduction, high-energy single-proton removal should, within the context of the
sudden approximation, leave the neutron configuration of the projectile unperturbed.  Here then, the
low-lying states populated in $^{16}$B should  
resemble a $\pi p_{3/2}$ hole coupled to the $^{17}$C
ground state neutron configuration.  Shell model calculations indicate that the latter is dominated 
($\sim$65 \%) by
approximately equal admixtures of $\nu (d_{5/2})^3_{J=3/2^+}$ and $\nu (d_{5/2}^2 s_{1/2})_{J=3/2^+}$  \cite{Mad01}, as confirmed by recent measurements of neutron removal from $^{17}$C \cite{Sau00,Mad01,Sau04}.  
As such a 0$^-$--~3$^-$ 
multiplet of states may be expected to be preferentially populated in $^{16}$B.

\begin{figure}[htb]
   \begin{center}
     \includegraphics[width=12cm]{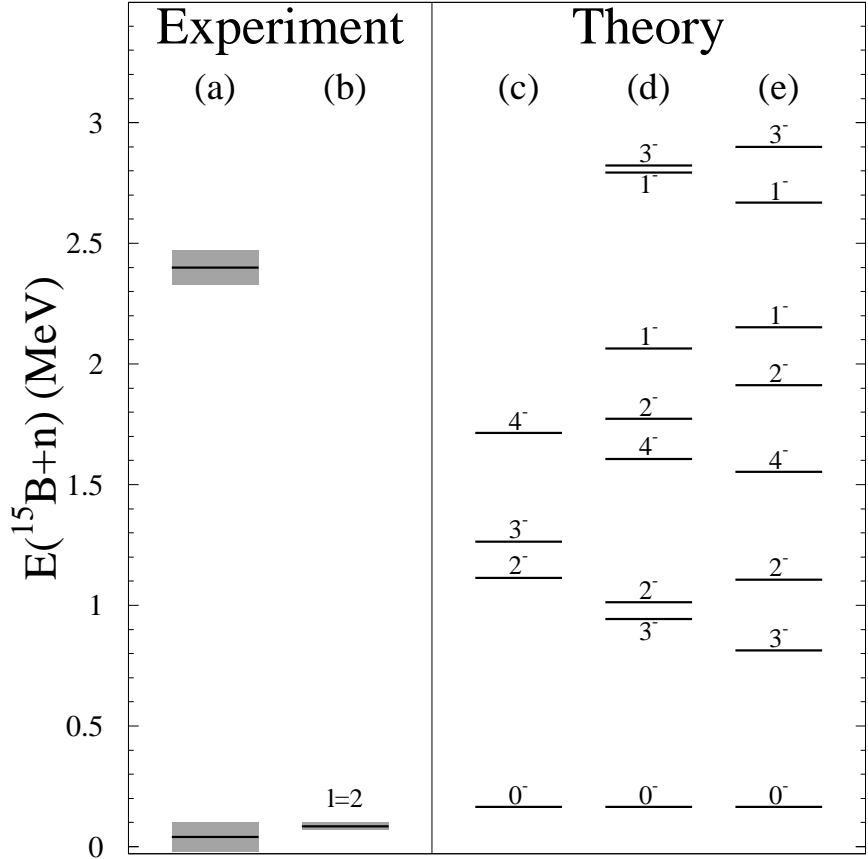}
   \end{center}
   \caption{\small \rm Compilation of energy levels in $^{16}$B measured in 
(a) the $^{14}$C($^{14}$C,$^{12}$N)$^{16}$B reaction study \cite{Kal00}, 
(b) the present work and those predicted below 3~MeV predicted by (0+1)$\hbar \omega $ shell model calculations:  
(c) Ref.~\cite{Pop85} (for which only the first four levels were tabulated), (d) Ref.~\cite{War92} and (e) the present work. The energy is given with respect to the neutron decay threshold. The shaded bands
correspond to the uncertainties in the measurements. In the case
of the shell model predictions, the energy of the ground state has been taken as the single-neutron separation
energy of 164~keV predicted by Ref.~\cite{War92}.}
   \label{fig:levels}
\end{figure}

Calculations carried out within the $p-sd$ model space using the WBP interaction \cite{War92}
predict that such a low-lying multiplet is indeed present in $^{16}$B (Figure \ref{fig:levels} and 
Table \ref{tab:sm}).  The states 
which are expected to be populated may be identified by the spectroscopic factors for
single-proton removal from the $\pi p_{3/2}$ orbital.  As listed in Table \ref{tab:sm}, the first 
four levels predicted to be strongly populated are the lowest lying 0$^-$--~3$^-$ states.
Energetically the first three members of this multiplet can only decay to the $^{15}$B ground state.
Of the other states calculated to be relatively strongly populated, only the 3$^-_2$ which 
is predicted to lie at 2.736~MeV could, in principle, produce a narrow low-lying line in the relative energy spectrum, such as that observed here, via
$d$-wave neutron decay to the second bound excited state of $^{15}$B (2.73~MeV, 7/2$^-$).  
However, only a small fraction of the decay is predicted to proceed in such a manner ($b_d$ = 0.04).
As such, it is reasonable to conclude that the peak observed in the \cbo\ relative-energy spectrum arises from one, or a combination, of the 0$^-_1$, 3$^-_1$ and 2$^-_1$ states.

The spectroscopic factors for the neutron decay
to the $^{15}$B ground state are listed in Table \ref{tab:sm}.
All the states predicted to be strongly populated 
are expected to decay essentially exclusively by $d$-wave neutron emission.  The single-particle
width for $d$-wave decay from a resonance at 100~keV is only 0.5~keV.  Clearly then,
the experimental resolution will dominate the lineshape of the low-lying states in the relative energy 
spectrum.  Assuming such a single, isolated low-lying resonance, described by an
$\ell$=2 Breit-Wigner lineshape modulated by the experimental
response function, and the uncorrelated \cbo\ distribution
described earlier, it was found that the reconstructed relative energy spectrum could be
very well reproduced for a resonance energy $E_r$=85$\pm$15~keV (Figure \ref{fig:narrow16b}).  Interestingly, this energy is compatible with that for the
lowest-lying feature observed in
the multi-nucleon transfer reaction study of Kalpakchieva 
et al. \cite{Kal00} (Figure \ref{fig:levels}).

\begin{figure}[htb]
   \begin{center}
     \includegraphics[width=12cm]{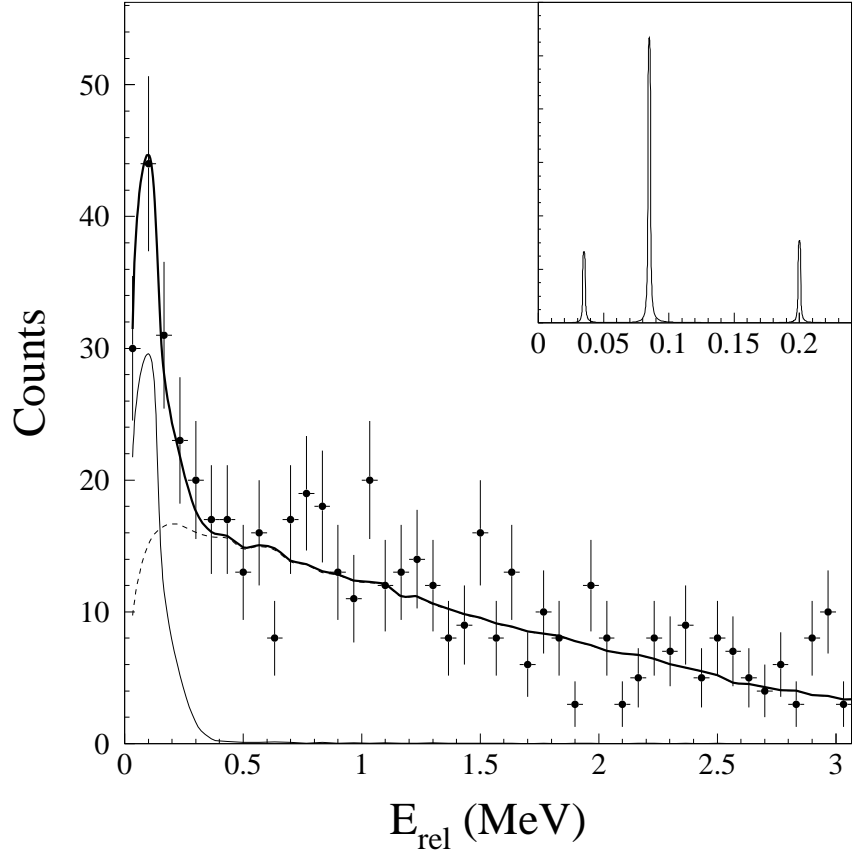}
   \end{center}
   \caption{\small \rm The \cbo\ relative energy spectrum compared to a simulation in which the separation
between the 0$^-_1$, 3$^-_1$ and 2$^-_1$ levels has been reduced (E$_r$=35, 85 and 200~keV) compared to the shell model calculations listed in Table \ref{tab:sm}.  The relative strengths correspond to those predicted (see inset).}
   \label{fig:fig5}
\end{figure}

Simulations undertaken assuming the population of the 0$^-_1$, 3$^-_1$ and 2$^-_1$ levels
according to the predicted energies and spectroscopic factors (Table \ref{tab:sm}), indicate 
that the 0$^-_1$ ground state would be observed as a narrow peak at threshold
well seperated from a higher lying and somewhat broader but more intense peak arising from 
the unresolved 3$^-_1$ and 2$^-_1$ levels \cite{Lec03}.
As displayed in Figure \ref{fig:fig5}, if the separation between the 0$^-_1$, 3$^-_1$ and 2$^-_1$ levels
is reduced to some 150~keV, rather than the predicted 950~keV, and the relative strengths remain as predicted, the simulated distribution 
matches the measured decay-energy spectrum.  
It has been noted previously that the shell model overestimates the excitation energies 
of bound states in this
mass region \cite{Sta04}, although to a much lesser degree than may be occuring here.  
This was attributed to the
weak binding of the valence neutrons.  It is conceivable, therefore, that such an
effect could be amplified in the case of an unbound system such as $^{16}$B.

As to the higher-lying states predicted to be
populated (E$_{rel} >$ 2~MeV), the degradation in the resolution with increasing
relative energy ({\sc fwhm} $>$ 0.45~MeV) makes their observation challenging.  Indeed, the most easily observable of these --
the 1$^-_1$ calculated to lie some 2~MeV above the ground state -- would, assuming the 
predicted strength and no feeding to the first excited state of $^{15}$B, be at the limits of detection given the present statistics.

In summary, the low-lying level structure of the unbound nucleus $^{16}$B has been investigated 
via single-proton removal from a $^{17}$C beam.  The reconstructed \cbo\ relative energy spectrum
exhibited a narrow resonance-like structure near threshold.  Simple 
considerations and comparison with shell-model calculations suggest that this peak arises
from a very narrow ($\Gamma \ll $100~keV) resonance, or an unresolved 
multiplet (0$^-_1$, 3$^-_1$, 2$^-_1$), with a
dominant $\pi (p_{3/2})^{-1} \otimes  \nu (d_{5/2}^3)_{J=3/2^+}$ + 
$\pi (p_{3/2})^{-1} \otimes  \nu (d_{5/2}^2,s_{1/2})_{J=3/2^+}$ configuration which decays 
by $d$-wave neutron emission.

Future measurements, with significantly improved resolution and higher statistics might
locate the
high-lying states and, moreover, attempt to resolve the possible multiplet
of near threshold states -- the latter being of some importance in the context of the
shell model predictions.  Whilst technically very challenging, the detection of gamma-rays in
coincidence with the \cbo\ fragments should be included to remove any possible ambiguity 
regarding the excitation energies of the levels in $^{16}$B.
Finally, it is interesting to note that the preliminary analysis of a measurement undertaken at RIKEN of 
neutron removal from 
$^{17}$B finds a \cbo\ relative energy spectrum almost identical to that reported here \cite{Nak06}.

\begin{acknowledgments}

The support provided by the technical staff of {\sc LPC} and the {\sc LISE} crew is gratefully acknowledged, 
as are the efforts of the {\sc GANIL} cyclotron operations team for providing the primary beam.
The assistance of S.M.G.~Chappell, S.M.~Singer, B. Benoit and L.~Donadille in preparing various elements of the experiment is also
acknowledged.  The theoretical advice on aspects of the structure of light, neutron-rich nuclei provided by D.J.~Millener is appreciated.
This work has been supported in part by the European Community within the FP6
contract EURONS RII3--CT-2004-06065.

\end{acknowledgments}

\begin{table}[htb]
   \begin{center}
       \begin{tabular}{ccccc}
        \hline \hline
         $J^\pi$ & $E_x$ (MeV) & $C^2S$ & $b_d$ & $b_s$  \\ \hline
          0$^-$   & 0.0   & 0.27  & 0.08 &       \\
          3$^-$   & 0.649 & 1.10  & 0.37 &       \\
          2$^-$   & 0.943 & 0.32  & 0.65 & 0.01  \\
          4$^-$   & 1.389 &       &      &       \\
          2$^-$   & 1.748 & 0.02  & 0.07 & 0.53  \\
          1$^-$   & 1.988 & 0.48  & 0.50 &       \\
          1$^-$   & 2.504 &       &      &       \\
          3$^-$   & 2.736 & 0.45  & 0.28 &       \\
          3$^-$   & 3.226 & 0.01  & 0.03 &       \\
          5$^-$   & 3.508 &       &      &       \\
          4$^-$   & 3.689 &       &      &       \\
          2$^-$   & 3.782 & 0.13  & 0.21 & 0.03  \\ 
          4$^-$   & 3.958 &       &      &       \\ \hline
       \hline
       \end{tabular}
       \caption{\small \rm Levels below 4.00~MeV in $^{16}$B predicted by (0+1)$\hbar \omega $ 
shell model calculations using 
the WBP interaction  \cite{War92} in the $p-sd$ valence space.
         $E_x$ is the excitation energy with respect to the 0$^-$
        ground state, which is calculated to lie 0.164~MeV 
above the \cbo\ threshold \cite{War92} ($E_x=E_r+0.164~MeV$); $C^2S$ is the spectroscopic 
factor for removing a 0$p_{3/2}$ proton from $^{17}$C;
and $b_d$ and $b_s$ are the spectroscopic factors for $d$ and $s$-wave neutron decay to
the $^{15}$B ground state (these are only listed for states with non-zero $C^2S$).}
      \label{tab:sm}
   \end{center} 
\end{table}

\begin{thebibliography}{99}
     \bibitem{Han95} see, for example, P.~G.~Hansen \etal\ , Ann. Rev. Nucl. Part. Sci. 45 (1995) 591
     \bibitem{Tho94} I.J.~Thompson and M.V.~Zhukov, \prc\ 49 (1994) 1904
     \bibitem{Tho96} I.J.~Thompson and M.V.~Zhukov, \prc\ 53 (1996) 708     
     \bibitem{Yam04} Y.~Yamaguchi \etal, \prc\ 70 (2004) 054320 and references therein
     \bibitem{Suz02} T.~Suzuki \etal, \prl\ 89 (2002) 012501
     \bibitem{Sta04} M.~Stanoiu \etal, \epja\ 22 (2004) 5
     \bibitem{Kan05} R.~Kanungo \etal, \plb\ 608 (2005) 206
     \bibitem{Kon05} Y.~Kondo \etal, \prc\ 71 (2002) 044611
     \bibitem{Kry96} R.~A.~Kryger \etal, \prc\ 53 (1996) 1971
     \bibitem{Kal00} R.~Kalpakchieva \etal, \epja\ 7 (2000) 451
     \bibitem{Zin95} M.~Zinser \etal, \prl\ 75 (1995) 1719
     \bibitem{Che01} L.~Chen \etal, \plb\ 505 (2001) 21
     \bibitem{Sau00} E.~Sauvan \etal, \plb\ 491 (2000) 1
     \bibitem{Mad01} V.~Maddalena \etal, \prc\ 63 (2001) 024613  
     \bibitem{Oga02} K.~Ogawa \etal, \epja\ 13 (2002) 81    
     \bibitem{Sau04} E.~Sauvan \etal, \prc\ 69 (2004) 044603
     \bibitem{Elk05} Z.~Elkes \etal, \plb\ 614 (2005) 174
     \bibitem{Ann87} R.~Anne \etal, \nima\ 257 (1987) 215
     \bibitem{Mar96} F.M.~Marqu\'es  \etal, \plb\ 381 (1996) 407  
     \bibitem{Til95} I.~Tilquin \etal, \nima\ 365 (1995) 446
	 \bibitem{Lec03} J.-L.~Lecouey, Thesis, Universit\'e de Caen (2002), LPCC T 02-03 \\ (http://tel.archives-ouvertes.fr/tel-00003117/fr/)
	 \bibitem{Gea87} R.~Brun \etal, GEANT 3 user's guide, CERN/DD/EE/84
     \bibitem{Lab99} M.~Labiche, Thesis, Universit\'e de Caen (1999), LPCC T 99-01     
     \bibitem{Lab01} M.~Labiche \etal, \prl\ 86 (2001) 600
     \bibitem{Lec07} J.-L.~Lecouey \etal, in preparation 
     \bibitem{Zaj84} W.~Zajc \etal, \prc\ 29 (1984) 2173
     \bibitem{Mar00} F.M.~Marqu\'es \etal, \plb\ 476 (2000) 219  
     \bibitem{War92} E.~K.~Warburton and B.~A.~Brown, \prc\ 46 (1992) 923
     \bibitem{Pop85} N.A.F.M.~Poppelier \etal, \plb\ 157 (1985) 120 
     \bibitem{Sta04} M.~Stanoiu \etal, \epja\ 22 (2004) 22
     \bibitem{Nak06} T.~Sugimoto \etal, RIKEN Accel. Prog. Rep. 37 (2004) 59
\end{thebibliography}
\end{document}